\documentclass{PoS}

\title{Development of a SiPM Camera for a Schwarzschild-Couder Cherenkov Telescope for the Cherenkov Telescope Array}

\ShortTitle{SCT-CTA Camera}

\author{\speaker{A.~N.~Otte}$^a$, J.~Biteau$^b$, H.~Dickinson$^c$, S.~Funk$^d$, T.~Jogler$^e$, C.~A.~Johnson$^b$, P.~Karn$^f$, K.~Meagher$^a$, H.~Naoya$^g$, T.~Nguyen$^a$, A.~Okumura$^g$, M.~Santander$^h$,  L.~Sapozhnikov$^e$, A.~Stier$^i$, H.~Tajima$^g$, L.~Tibaldo$^e$ ,  J.~Vandenbroucke$^f$, S.~Wakely$^i$, A.~Weinstein$^c$,  D.~A.~Williams$^b$, for the CTA Consortium$^j$ \\
\llap{$^a$} School of Physics \& Center for Relativistic Astrophysics, Georgia Institute of Technology, 837 State Street NW, Atlanta, GA 30332-0430, USA\\
\llap{$^b$} SCIPP and Department of Physics, University of California, Santa Cruz, CA  95064, USA\\
\llap{$^c$} Physics \& Astronomy, Iowa State University, Ames, IA 50011-3160, USA\\
\llap{$^d$} Erlangen Centre for Astroparticle Physics, Friedrich-Alexander-Universit\"at Erlangen-N\"urnberg, Erwin-Rommel-Str. 1, D-91058 Erlangen, Germany\\
\llap{$^e$} Kavli Institute for Particle Astrophysics and Cosmology, SLAC National Accelerator Laboratory, Stanford University, Stanford, CA 94305, USA\\
\llap{$^f$} Department of Physics and Wisconsin IceCube Particle Astrophysics Center, University of Wisconsin, Madison, WI 53706, USA\\
\llap{$^g$} Solar-Terrestrial Environment Laboratory, Nagoya University, Furo-cho, Chikusa-ku, Nagoya, Aichi 464-8601, Japan\\
\llap{$^h$} Department of Physics and Astronomy Barnard College, Columbia University, New York, USA\\
\llap{$^i$} Enrico Fermi Institute, University of Chicago, 933 E 56th St, Chicago, IL 60637, USA\\
\llap{$^j$} Full consortium author list at http://cta-observatory.org\\
E-mail: \email{otte@gatech.edu}}

\abstract{We present the development of a novel 11328 pixel silicon photomultiplier (SiPM) camera for use with a ground-based Cherenkov telescope with Schwarzschild-Couder optics as a possible medium-sized telescope for the Cherenkov Telescope Array (CTA). The finely pixelated camera samples air-shower images with more than twice the  optical resolution of cameras that are used in current Cherenkov telescopes. Advantages of the higher resolution will be a better event reconstruction yielding improved background suppression and angular resolution of the reconstructed gamma-ray events, which is crucial in morphology studies of, for example, Galactic particle accelerators and the search for gamma-ray halos around extragalactic sources.  Packing such a large number of pixels into an area of only half a square meter and having a fast readout directly attached to the back of the sensors is a challenging task. For the prototype camera development, SiPMs from Hamamatsu with through silicon via (TSV) technology are used. We give a status report of the camera design and highlight a number of technological advancements that made this development possible.}

\FullConference{The 34th International Cosmic Ray Conference,\\
		30 July- 6 August, 2015\\
		The Hague, The Netherlands}

\begin{document}

\begin{figure}[t]
        \centering
   \includegraphics*[angle=0,width=0.8\textwidth]{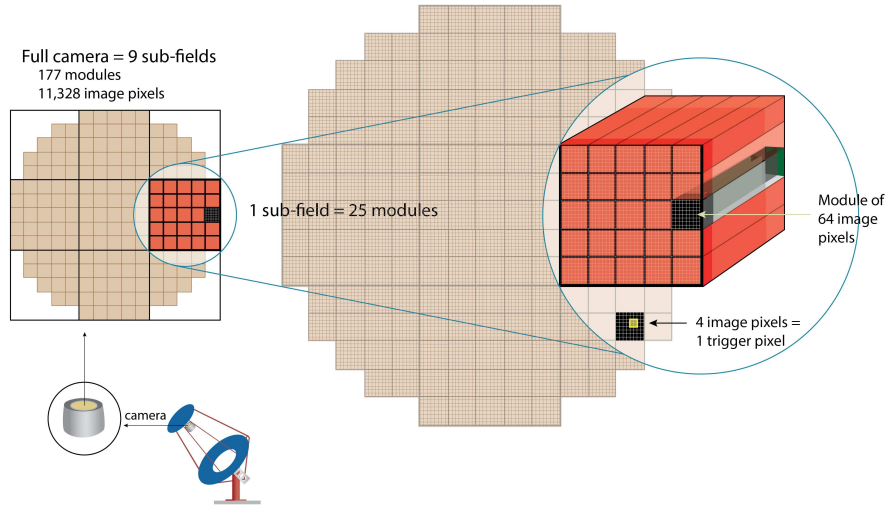}
   \caption{\small Conceptual layout of the camera focal plane
                        \label{topology}}
\end{figure}

\section{Introduction}

Exploring the universe in the very-high energy (VHE) gamma-ray band has resulted in a plethora of stunning discoveries and many paradigm shifts in past years. From observations with the latest generation of Cherenkov telescope arrays it has come clear that our view of the VHE universe is limited by the sensitivity of existing instruments. These limitations spurred the VHE community to pursue the construction of the Cherenkov Telescope Array (CTA) with the goal to deliver a VHE instrument with one-order-of-magnitude improved sensitivity. A detailed description of CTA and its science goals can be found in \cite{2013APh....43....3A}. 

The CTA group in the United States has proposed an innovative, two-mirror Schwarzschild-Couder telescope (SCT) \cite{2007APh....28...10V} array for CTA. The demagnifying SCT optical design provides a dramatic reduction in the plate scale compared to currently used telescopes, enabling a camera based on modern technology including self-triggering custom ASICs with high channel density as well as SiPMs.  We have designed a low-cost camera based on SiPM photon detectors, the TARGET custom analog pipeline ASICs, a backplane to merge a large number of channels into a single high-speed FPGA for the pattern trigger, and another FPGA for data acquisition.

Section \ref{camera} gives an overview of the SCT camera. Section \ref{SiPMs} describes the photon detectors. How they are being integrated into the camera is being detailed in Section \ref{integration}. In Section \ref{electronics}, a description of the front-end electronics and data acquisition is given. The paper closes with an outlook of the next steps in camera production in Section \ref{outlook}.  

\begin{figure}[t]
        \centering
\includegraphics*[angle=0,width=0.49\textwidth]{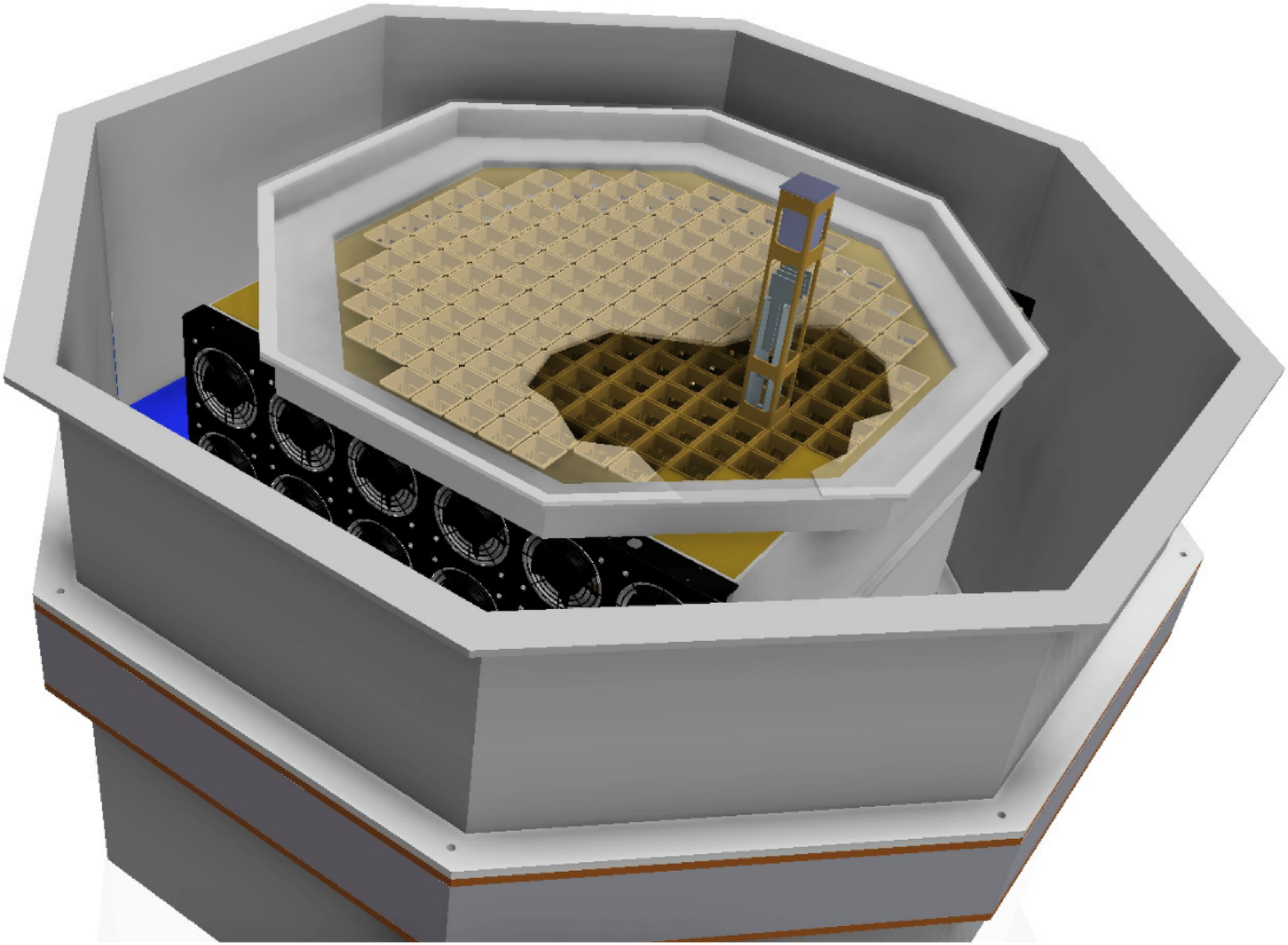}
\includegraphics*[angle=0,width=0.49\textwidth]{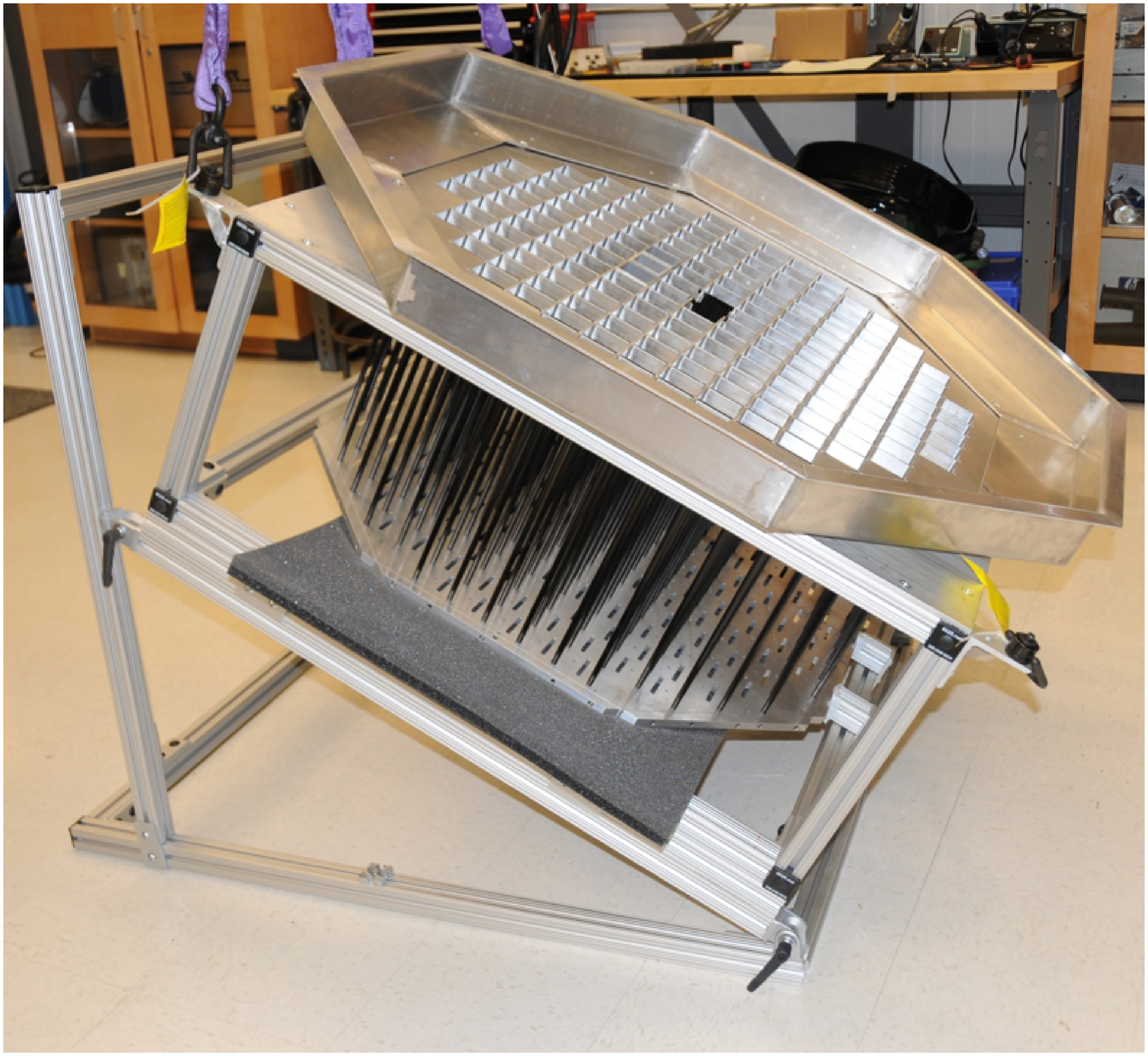}
                        \caption{\small The left image is a model of the camera and illustrates how one camera module is inserted into the camera. The image on the right shows parts of the fabricated inner camera. Shown are the front lattice and the focal plane enclosure mounted on top of the front lattice. Behind the front lattice can be seen the carbon fibre rods of the drawer structure and the bulk plate.
						\label{SCTcamera}}
\end{figure}

\section{Layout of the camera }\label{camera}

The three main design drivers of the camera are: (1) A hierarchical architecture that is easily serviceable and contains few failure points in the form of cable interconnects, (2) precise mechanical registry and adjustment of the SiPMs and (3) thermal/environmental control to remove waste heat from the densely packed camera electronics while providing temperature regulation for the photosensors.  

The camera consists of an inner camera and an outer enclosure. The outer enclosure acts as a shield from the elements and serves as interface to the telescope structure. The inner camera hosts the photosensors, front-end electronics, digitization electronics, level-0 (L0) trigger, and back-plane electronics that provide the pattern trigger and data acquisition hardware. 

\begin{figure}[t]
        \centering
   \includegraphics*[angle=0,width=\textwidth]{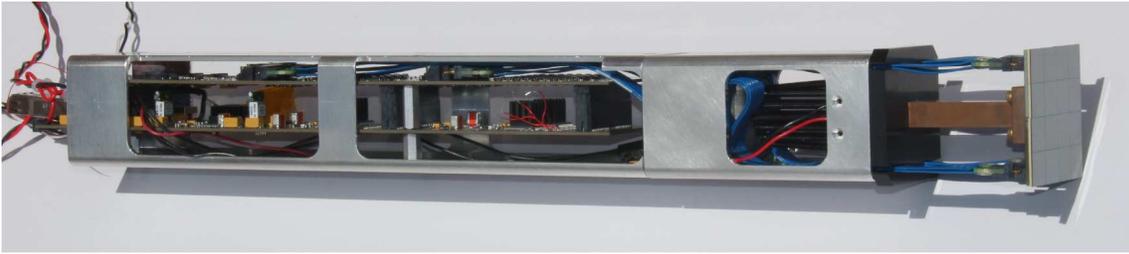}
   \caption{\small Picture of one fully assembled camera module used for prototype testing. See text for details.
                        \label{assembledmodule}}
\end{figure}

The camera architecture is subdivided into nine subfields or sectors (see Figure \ref{topology}). A fully loaded sector has 25 modules that are inserted from the front into the drawer-like structure of the inner camera (see left panel in Figure \ref{SCTcamera}). A module consists of 64 imaging pixels in an eight-by-eight  configuration. The size of one pixel is $6.5\times6.5$\,mm$^2$, which corresponds to an angular size of $0.067^\circ$ in the sky. Integrated in each module is the complete readout electronics (see Figure \ref{assembledmodule}). A fully equipped camera has 177 modules with a total of 11,328 pixels.

The SC optics requires positional uncertainties of the focal plane along the optical axis that are less than $200\,\mu$m.  The primary positional reference for the focal plane is provided by a 3/4" machined aluminum lattice plate (see right panel in Figure \ref{SCTcamera}) that is mounted to a rack of thermally-stable carbon-fiber supports. The modules slide into the front lattice and ultimately register on its front surface through a Delrin collar. Relative to the front lattice, the photon detectors of each module are adjusted to a different height to approximate the ideal curvature of the SC optics. The modules are locked in place by a pulling mechanism that acts on the module from the back of the camera. Deformations between the front lattice and back of the camera due to temperature changes are compensated by dividing the module frame into two separated pieces that can move relative to each other.

Dissipating the about 3\,kW of heat produced by the camera electronics is accomplished with forced cooled air. Thermally insulated from the camera electronics are the photon detectors that are encapsulated in the so-called focalplane box and isolated from the environment with a UV transparent acrylic window.

\section{Photon detector used in the prototype SCT \label{SiPMs}}

\begin{figure}[t]
        \centering
\includegraphics*[angle=0,width=5cm]{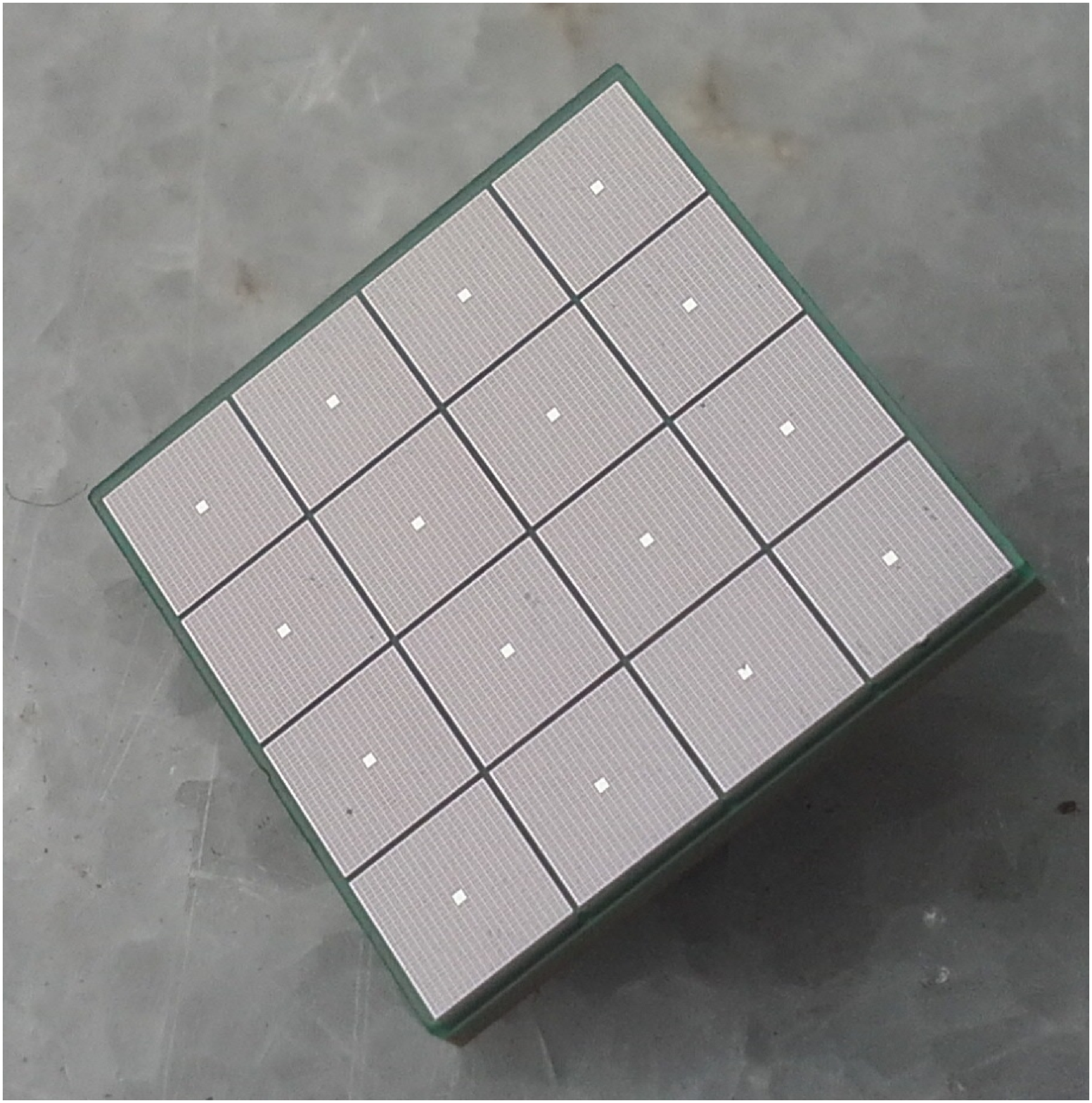}
\includegraphics*[angle=0,width=9cm]{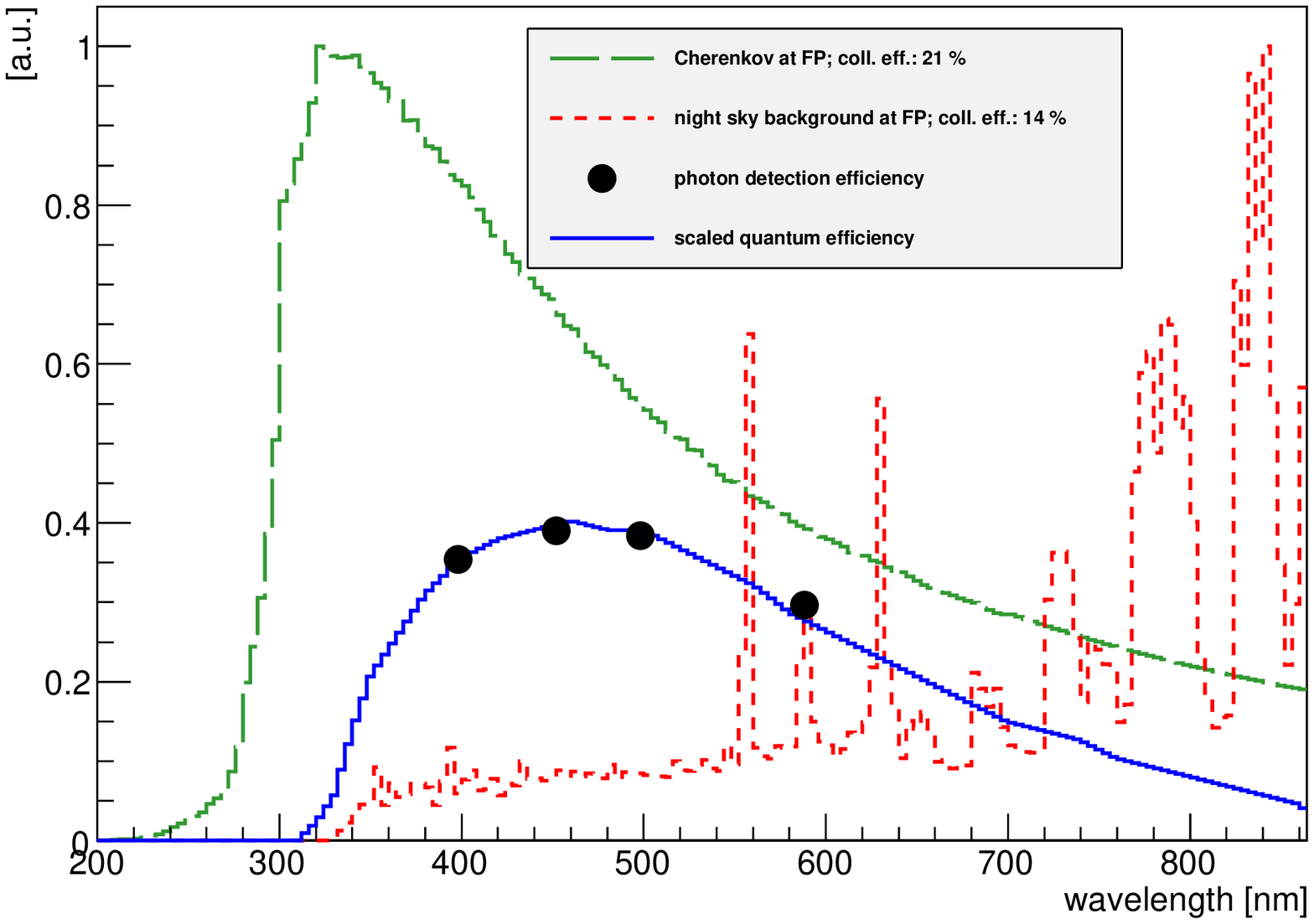}
                        \caption{\small Left:  One tile of the Hamamatsu S12642-0404PA-50(X) used in the SCT prototype. Right: Photon detection efficiency of the Hamamatsu S12642-0404PA-50(X) when operating at 3\,V overvoltage (black) and quantum efficiency (blue) scaled to match the photon detection efficiency (PDE). For the efficiency curves, the absolute efficiency can be read off the labeled y-axis. The green curve gives the Cherenkov spectrum in the focal plane and the red curve the night sky background spectrum in the focal plane. Integrating over both spectra the S12642 detects 21\% of the Cherenkov photons that arrive in the focal plane and 14\% of the night sky background when operated at 3\,V overvoltage. 
						\label{PDE}
  }

\end{figure}

A large number of silicon photomultipliers (SiPMs) have been evaluated from different vendors \cite{2013SPIE.8852E..0KB} and the Hamamatsu S12642-0404PA-50(X) was selected as the photon detector for the SCT prototype. The S12642 is a tile of 16 SiPMs, each $3\times3$\,mm$^2$ in size (see Figure \ref{PDE}). The microcell pitch of the SiPMs is 50\,$\mu$m. In the prototype SCT, groups of 4 SiPMs are connected in parallel to form one $6\times6$\,mm$^2$ pixel in the readout. One main criterion to choose the S12642 over other devices is its use of the through silicon via (TSV) technology, which minimizes dead-space at the boundary of the pixel. 

Extensive tests have been performed on individual tiles, as well as basic functionality tests on 415 tiles that have been purchased for the 25 modules of the prototype SCT. Figure \ref{PDE} shows the photon detection efficiency of one SiPM measured at 3\,V overvoltage. Overlaid over the photon detection efficiency (PDE) measurements is the quantum efficiency (QE) curve measured by Hamamatsu after scaling the QE curve to match the PDE measurements.

 Optical crosstalk is fairly high with 50\% at an overvoltage of 3\,V (5\% above breakdown), which raises the trigger threshold of the SCT. The spectral response of the S12642 is such that 21\% of the Cherenkov photons are detected and 14\% of the unwanted night sky background photons. Newer devices that have become available from several vendors lately have much lower optical crosstalk and a spectral response that is more favorable for detecting Cherenkov light and will be considered in the future \cite{biteau}.
 
\section{Mechanical integration and temperature stabilization of the SiPMs\label{integration}}

The photon detectors for one camera module are divided into four quadrants. Each quadrant has four SiPM tiles with 16 image pixels. Four quadrants are assembled into a focal-plane module, which in turn is attached to the top of a camera module. The sensors are thermally insulated from the back of the camera by a plastic base plate and one inch thick insulation foam (densified Solemide from Evonik). Active cooling of the sensors is accomplished with a thermoelectric element that is thermally coupled to a copper post. Temperature feedback is provided by a thermistor that is mounted on the back side of each quadrant. When the telescope is in operation, the temperature will be stabilized to a temperature a few degrees below the equilibrium temperature of the focal plane box to minimize the required cooling power. See \cite{NDIPProceeding} for further details on the photon detector integration.

\section{Readout electronics \label{electronics}}

The complete readout electronics is integrated inside the camera. Front-end signal processing, signal digitization, and the first level trigger take place inside the camera modules. The electronics inside the modules also includes ancillary systems such as current monitors for tracking bright stars, high-voltage distribution to the photon detectors, and temperature control for the photon detectors. 

Communication between the camera modules and the camera servers, distribution of trigger signals, power distribution, et cetera are managed by camera backplanes, one for each camera sector. The data from the nine camera sector backplanes are merged and delivered to a level-2 event-building system (the SCT Camera Server) using standard high-speed communication protocols (including 10 Gb/s-Ethernet switches). Trigger and time synchronization signals from the backplanes are relayed by the distributed array trigger electronics. 
The main camera electronic components are described in the following.

\subsection{Preamplifier}
The raw signals from the S12642 MPPCs exhibit a fast rising edge and a slow decaying tail ($>100$\,ns) due to the large terminal capacitance ($>1$\,nF) of the four, connected in parallel SiPMs to form one pixel. For optimum signal separation from fluctuations in the night-sky-background, a full width at half maximum (FWHM) of 10\,ns or less is desirable for the SiPM signals. A shorter pulse in the readout is achieved by differentiating the signals with a high-pass filter with pole zero cancellation in between the two stages of the preamplifier (see \cite{NDIPProceeding} for further details). The resulting FWHM of 10\,ns of the output signal is dominated by the limited bandwidth of the Analog Devices AD8014 chip that is used to amplify the signal. While a higher bandwidth would be advantageous, the two stage amplifier consumes only 10\,mW, which amounts to $\sim100$\,W of waste heat for the 11,328 pixel camera.  The dynamic range of the preamplifier is 250 photoelectrons. Further developments of preamplifier concepts that aim at faster signals and less power consumption are ongoing and will be employed in the final SCT array. The amplitude of a single photoelectron signal at the output of the amplifier is 4\,mV at a 50\,$\Omega$ impedance for an operating voltage of the SiPM that is 3\,V above breakdown.

\subsection{Signal digitization and first level trigger}

\begin{figure}[t]
        \centering
   \includegraphics*[angle=0,width=0.40\textwidth]{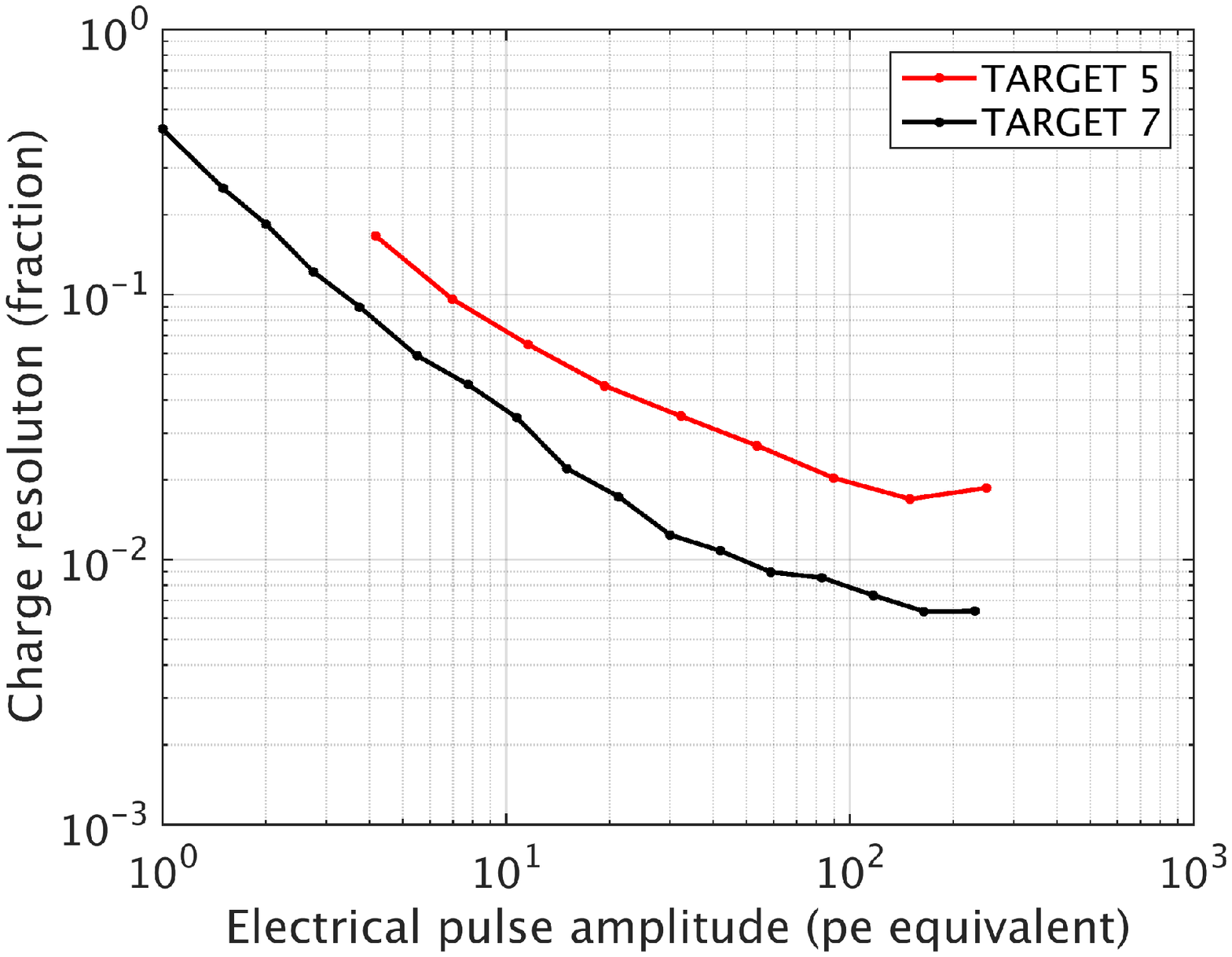}
   \includegraphics*[angle=0,width=0.49\textwidth]{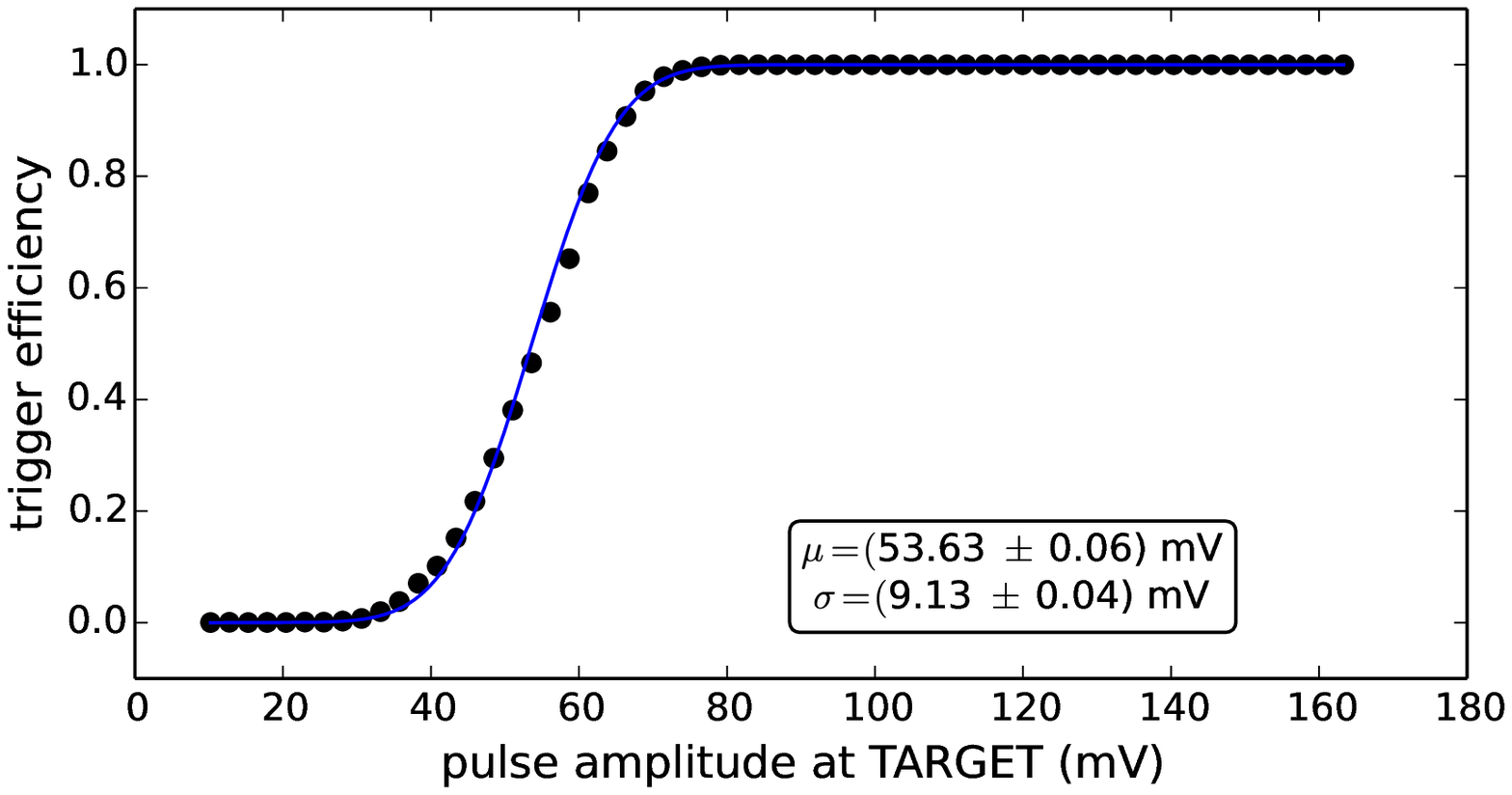}
   \caption{\small Performance of TARGET~7 evaluated using signals from a function generator. Left: charge resolution as a function of input charge (including the contribution from the ASIC only) assuming the 4\,mV/p.e. pulse amplitude expected in the prototype SCT camera. Right: demonstration of the minimum stable trigger threshold achievable by TARGET~7 during normal operations. The plot shows the trigger efficiency as a function of input pulse amplitude with best fit Gaussian error function: the inset panel gives the Gaussian mean (threshold) and standard deviation (noise).
                        \label{T7performance}}
\end{figure}

After signal amplification and shaping, digitization and the level-1 trigger are integrated in an application-specific integrated circuit (ASIC) TARGET \cite{2012APh....36..156B}. The prototype SCT uses the $7^{th}$ generation TARGET (TARGET~7). A TARGET chip has 16 input channels, each equipped with a 16,384 cell switched capacitor array (SCA) providing a buffer depth of $16.384\,\mu$s at a sampling rate of 1\,GSa/s. When a readout command is issued, an on-chip 10-bit (effective) Wilkinson analog-to-digital converter digitizes the sampled voltage levels stored in the SCA. TARGET 7 has a dynamic range of 1.9\,V. Figure~\ref{T7performance} shows the charge resolution of the TARGET-7 chip.

In parallel to signal sampling and digitization, TARGET also performs the first stage of the trigger by forming the analog sum of four adjacent image pixels and discriminating the summed signal. A trigger efficiency curve as a function of input pulse amplitude evaluated in a bench test using signals from a function generator is shown in Figure \ref{T7performance}. We note that the minimum stable trigger threshold during normal operations for TARGET~7 is 50\,mV and the typical trigger noise is 10--15\,mV due to coupling between the sampling and triggering circuits of the ASIC. A newer generation of TARGET ASICs, currently undergoing testing, is expected to circumvent the coupling by separating the sampling/digitization and triggering into two separate ASICs, TARGET~C and CCTV. Based on the experience with previous TARGET generations we expect to achieve a minimum trigger threshold of 4.5\,mV and a trigger noise of $\sim 0.5$\,mV. More information about TARGET and details about the performance of TARGET 7 can be found in \cite{TARGETProcPaper}.

\subsection{Camera backplane}

The (up to) 400 trigger signals in each camera subfield are presented to the level-1 (L1) trigger FPGA on the backplane where the relative timing is adjusted to ns precision and patterns of adjacent hits are located with an essentially dead-time-free pipelined architecture with a coincidence resolving time programmable down to 4 ns. The array trigger and backplane electronics provide phase-locking of the 125 MHz clock through the entire camera electronics to the individual TARGET ASICs, with synchronization good to 1 ns. 

The level-2 (L2) trigger collects L1 triggers from the backplane and provides other timing signals required for event synchronization, event tagging, clock synchronization, and synchronous resets. Each backplane takes the 1600 1-GSa/s waveform data streams from 25 camera modules and combines them in a data acquisition FPGA, where buffering and event building (and, optionally, zero suppression, trace integration, and/or data compression) are performed before outputting the data on a pair of 10 Gb/s Ethernet links. Data are merged with a high-speed switch so that the backplanes can serve data and command telemetry to a number of client computers. One client is the Camera Server computer, which includes the level-2 event builder (combining all data from the multiple backplanes) and an interface to the central DAQ.

The data acquisition electronics are designed to handle a single telescope trigger rate up to 10 kHz. However, a suppression factor of at least 4 is provided by the distributed trigger array system, leading to an effective single-telescope trigger rate of 2.5 kHz. For a full-camera readout size of 1.42 MB (125 bytes/pixel for 11,328 pixels) this trigger rate translates into a raw data rate of 3.54 GB/s (28.3 Gb/s) per telescope. If additional reduction is needed, the backplane FPGAs can perform data compression, zero suppression, and/or waveform integration prior to transmission. 

\subsection{Distributed Intelligent Array Trigger}

The Distributed Intelligent Array Trigger (DIAT) assumes responsibility for the level-2 (L2) trigger event time stamps (camera trigger), and synchronized clock distribution in the camera with nanosecond precision.

DIAT will enable computation of low-order moments of the shower image, which are subsequently sent to the DIAT modules of neighbouring SCTs. In addition to a more traditional telescope multiplicity trigger, application of a hardware-level algorithm that combines image-moment data from nearby SCTs to compute a real-time localized level-3 (L3) trigger decision is also an option \cite{DIATThisProceeding}. Both allow spurious L2 triggers at a single telescope to be identified and vetoed before camera readout, ameliorating the data-transfer rates for the SCT subarray.

A prototypical trigger demonstrator has been fabricated, which will receive and temporarily store timestamped trigger messages from the local backplane modules. Those messages can be accessed and analyzed asynchronously using slow-control software. It will use a special calibration mode to compute the signal propagation delays that can be used to calibrate the embedded clock signal that it will transmit to each local backplane.

\section{Outlook\label{outlook}}

The construction of the camera for the prototype SCT is ongoing. For the prototype SCT one camera sector will be equipped with camera modules. The photon detectors have been tested and are now being assembled into focal plane modules. In parallel, prototypes of the front-end electronic boards have been tested and production of boards for 29 camera modules is ongoing. The fabrication of the camera mechanics, the backplane for one sector and the DIAT are also underway. After integration and final testing of the camera is complete, the camera will be shipped to the VERITAS site for assembly onto the SCT, which is expected to take place in Fall 2015.

\section{Acknowledgements}
We gratefully acknowledge support from the agencies and organisations listed in this page: http://www.cta-observatory.org/?q=node/22. The development of the prototype SCT has been made possible by funding provided through the U.\ S.\ NSF-MRI program. 

\end{document}